\documentclass[11pt]{article}

\usepackage[T1]{fontenc}
\usepackage{newtxtext,newtxmath}
\usepackage[scaled=0.85]{beramono}
\usepackage{geometry}
\usepackage{setspace}
\usepackage[
backend=biber,
style=numeric,
sorting=none,
maxnames=8,        
minnames=1,        
giveninits=true,   
]{biblatex}

\renewbibmacro{in:}{%
  \ifentrytype{article}
    {}
    {\printtext{\bibstring{in}\intitlepunct}}%
}
\AtEveryBibitem{
  \clearfield{urlyear}
  \clearfield{urlmonth}
  \clearfield{urlday}
}

\usepackage{graphicx}
\usepackage{caption}
\usepackage{mathtools}
\usepackage{bm}

\usepackage{amsthm}
\usepackage{siunitx}
\usepackage{booktabs}
\usepackage{dirtree}
\usepackage{multicol}
\usepackage{orcidlink}

\usepackage{listings}
\usepackage{xcolor}

\definecolor{bg}{rgb}{0.97,0.97,0.97}
\definecolor{comment}{rgb}{0.25,0.5,0.35}
\definecolor{keyword}{rgb}{0.0,0.2,0.6}
\definecolor{string}{rgb}{0.58,0.0,0.82}
\definecolor{number}{rgb}{0.5,0.5,0.5}
\definecolor{frame}{rgb}{0.8,0.8,0.8}

\lstdefinestyle{mintedstyle}{
    backgroundcolor=\color{bg},
    basicstyle=\ttfamily\scriptsize,
    breaklines=true,
    breakatwhitespace=true,
    columns=fullflexible,
    keepspaces=true,
    showstringspaces=false,
    tabsize=4,
    frame=none,
    rulecolor=\color{frame},
    framesep=2mm,
    numbers=left,
    numberstyle=\tiny\color{number},
    numbersep=8pt,
    xleftmargin=2em,        
    keywordstyle=\color{keyword}\bfseries,
    commentstyle=\color{comment}\itshape,
    stringstyle=\color{string},
    captionpos=b
}

\usepackage{tikz}
\usetikzlibrary{positioning, arrows.meta, shapes.geometric}
\usepackage{hyperref}
\hypersetup{
    colorlinks=true,
    linkcolor=blue,
    urlcolor=blue,
    citecolor=blue,
    pdfborder={0 0 0}
}
\makeatletter
\def\Hy@raisedlink#1{#1}  
\makeatother

\title{\Large\bfseries
Basilisk and Docker for Reproducible GN\&C Simulation: A Workflow Reference
}

\let\oldeqref\eqref
\renewcommand{\eqref}[1]{Eq.~\oldeqref{#1}}

\author{
\large Anubhav Gupta\thanks{Email: \texttt{anubhav.gupta@colorado.edu}}\;
\orcidlink{0000-0002-3216-868X}\\[4pt]
\small University of Colorado Boulder, CO 80303, USA\\
\small In Orbit Aerospace Inc., Torrance, CA 90501, USA
}

\date{}

\begin{document}
\maketitle
\vspace{-0.75em}

\begin{abstract}
Basilisk is an open-source astrodynamics simulation framework widely used for spacecraft guidance, navigation, and control (GN\&C) research and development. Despite its flexibility and computational capabilities, configuring Basilisk consistently across heterogeneous development environments presents practical challenges due to dependency management, operating system compatibility, 
and software configuration requirements. This paper presents a Docker-based containerization workflow for Basilisk that encapsulates the complete build environment, dependencies, and simulation infrastructure within a portable container image. The workflow is demonstrated 
through a progression of simulation scenarios of increasing complexity, from standalone orbital dynamics scripts to BSKSim-based attitude dynamics and control simulations with Monte Carlo analysis. The BSKSim class hierarchy, dynamics model architecture, flight software implementation, and scenario execution patterns are described in detail. The presented workflow provides a self-contained implementation reference for GN\&C engineers and researchers seeking reproducible and 
portable Basilisk simulation environments. This work expands upon a workshop presentation delivered at the 46th Rocky Mountain AAS GN\&C Conference, February 2024, available at \url{https://doi.org/10.5281/zenodo.15008785}.
\end{abstract}


\section*{Nomenclature}

\begin{multicols}{2}
{\renewcommand\arraystretch{0.92}
\noindent\begin{tabular}{@{}ll@{}}
$n$             & mean orbit rate, s$^{-1}$ \\
$\mu$           & gravitational parameter, km$^3$/s$^2$ \\
$a$             & semi-major axis, km \\
$e$             & orbital eccentricity \\
$i$             & inclination, rad \\
$\Omega$        & RAAN, rad \\
$\omega$        & argument of periapsis, rad \\
$f$             & true anomaly, rad \\
$T$             & orbital period, s \\
$T_{\text{final}}$ & total simulation time, s \\
\noalign{\smallskip}\hline\noalign{\smallskip}
\multicolumn{2}{@{}l}{\textit{Subscripts}} \\
$J_2$   & $J_2$ perturbation \\
$BN$    & body / inertial frame \\
$CN$    & CoM / inertial frame \\
\end{tabular}}

\columnbreak

{\renewcommand\arraystretch{0.92}
\noindent\begin{tabular}{@{}ll@{}}
$\mathbf{r}$         & position vector, km \\
$\hat{\mathbf{r}}$   & unit position vector \\
$\ddot{\mathbf{r}}$  & acceleration vector, km/s$^2$ \\
$\mathbf{a}_{J_2}$   & $J_2$ acceleration, km/s$^2$ \\
$\mathbf{I}_{sc}$    & inertia matrix, kg$\cdot$m$^2$ \\
$K, P$               & MRP control gains \\
$N$                        & number of logged data points \\
$\Delta t_\text{sim}$      & simulation time step, s \\
$\Delta t_\text{samp}$     & sampling period, s \\
\end{tabular}}
\end{multicols}

\section{Introduction}
\label{sec:intro}
Spacecraft guidance, navigation, and control (GN\&C) simulations play a critical role in the design, validation, and analysis of modern space missions, where even minor malfunctions may contribute to mission failure~\cite{kenneally2020basilisk}. Such simulations are frequently used to evaluate spacecraft dynamics, flight software behavior, mission operations concepts, estimation algorithms, and autonomous control strategies prior to deployment. As spacecraft missions increase in complexity, the supporting simulation environments likewise require increasingly sophisticated software infrastructures capable of integrating dynamics models, visualization tools, numerical solvers, and custom flight software components. Existing astrodynamics and spacecraft simulation tools span commercial off-the-shelf, government-developed, and open-source ecosystems, including platforms such as STK, MATLAB/Simulink, GMAT, NASA 42, NASA Trick, and DARTS/DShell~\cite{kenneally2020basilisk}.

Basilisk (BSK) is an open-source astrodynamics simulation framework developed collaboratively by the Autonomous Vehicle Systems (AVS) Laboratory at the University of Colorado Boulder and the Laboratory for Atmospheric and Space Physics (LASP)~\cite{kenneally2020basilisk}. The framework provides a modular simulation architecture for spacecraft dynamics and flight software development while supporting both faster-than-real-time and real-time simulation workflows. Its underlying C++ implementation delivers high-speed execution, while a Python interface ensures accessibility for GN\&C engineers and researchers. Basilisk has been applied to a broad range of applications including astrodynamics research, spacecraft autonomy studies, mission concept development, and hardware-in-the-loop simulation environments. Basilisk also integrates with Vizard~\cite{wood2018flexible}, a Unity-based three-dimensional visualization tool capable of rendering spacecraft states and simulation telemetry interactively.

Despite the flexibility provided by modern simulation frameworks, configuring portable and repeatable software environments for astrodynamics simulations can remain challenging. Scientific software stacks often depend on operating system compatibility, compiler toolchains, Python environments, Conan package management, external libraries, visualization dependencies, and ephemeris datasets. Variations across development machines may therefore introduce inconsistencies in installation procedures, software behavior, or simulation execution workflows.

Containerization technologies such as Docker~\cite{merkel2014docker} provide one possible solution to these challenges by packaging software applications and their dependencies into isolated and portable runtime environments. Such workflows simplify software deployment, reduce dependency conflicts, and improve portability across development systems~\cite{rad2017introduction}. In the context of GN\&C simulation workflows, containerized environments can help streamline installation procedures and support consistent execution environments for research, collaboration, and educational activities.

This work expands upon the workshop presentation titled ``Basilisk and Docker for Streamlined GN\&C Simulation,'' presented at the 46th Rocky Mountain AAS GN\&C Conference in February 2024~\cite{gupta2024basilisk}, openly available through Zenodo at \url{https://doi.org/10.5281/zenodo.15008785}. The purpose of this document is to provide an expanded implementation reference covering Docker-based Basilisk deployment, modular simulation architecture, scenario-based scripting workflows, BSKSim-based simulation frameworks, visualization integration, and simulation data recording utilities.

The remainder of this paper is organized as follows. Section~\ref{sec:background} provides background on Basilisk, Vizard, and Docker. Section~\ref{sec:containerization} describes the Docker configuration for Basilisk. Section~\ref{sec:architecture} covers Basilisk's software architecture. Section~\ref{sec:standalone} presents standalone-script simulation workflows. Section~\ref{sec:bsksim} presents BSKSim-based simulation workflows including attitude dynamics, control, and Monte Carlo analysis. Section~\ref{sec:discussion} discusses practical implications and limitations, and Section~\ref{sec:conclusion} concludes the paper.

\section{Background}
\label{sec:background}

This section provides an overview of the three core components 
used in this work: the Basilisk astrodynamics simulation 
framework, the Vizard visualization tool, and the Docker 
containerization platform.

\subsection{Basilisk}
\begin{figure}[htbp]
\centering
\begin{tikzpicture}[
    node distance=1cm,
    every node/.style={font=\small},
    block/.style={
        rectangle, rounded corners, draw=black, thick,
        minimum width=3.2cm, minimum height=0.8cm, align=center
    },
    arrow/.style={-{Latex[length=2.5mm]}, thick}
]
\node[block] (sim) {Simulation Container};
\node[block, below=of sim] (proc) {Processes};
\node[block, below=of proc] (task) {Tasks};
\node[block, below left=0.9cm and 1.6cm of task]  (mod1) {Dynamics Module};
\node[block, below=0.9cm of task]                 (mod2) {FSW Module};
\node[block, below right=0.9cm and 1.6cm of task] (mod3) {Sensor Module};
\node[
    rectangle, dashed, draw=black,
    minimum width=10cm, minimum height=2.2cm,
    below=0.2cm of task,
    label={[yshift=-2.1cm]\small Message-Passing Architecture}
] (msgbox) {};
\draw[arrow] (sim) -- (proc);
\draw[arrow] (proc) -- (task);
\draw[arrow] (task) -- (mod1);
\draw[arrow] (task) -- (mod2);
\draw[arrow] (task) -- (mod3);
\draw[arrow,<->] (mod1) -- (mod2);
\draw[arrow,<->] (mod2) -- (mod3);
\end{tikzpicture}
\caption{Simplified Basilisk execution hierarchy and 
         module-message architecture.}
\label{fig:basilisk_architecture}
\end{figure}
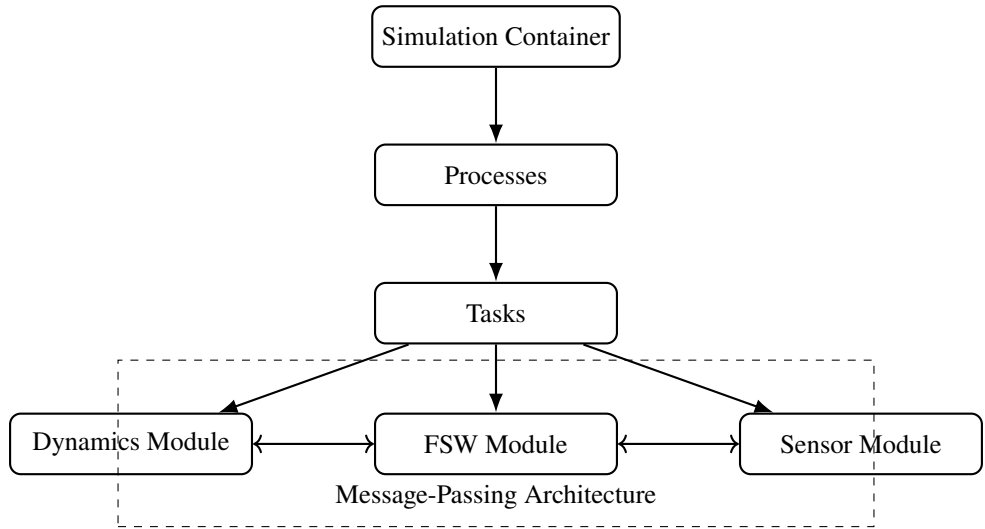

Basilisk (BSK) is an open-source astrodynamics simulation framework that supports spacecraft dynamics, guidance, navigation, and control (GN\&C) simulations through a modular architecture composed of interconnected software modules and message-passing interfaces. Simulation functionality is organized through reusable modules representing spacecraft dynamics, sensors, actuators, estimation routines, and flight software components, which communicate through structured message interfaces enabling flexible construction of mission-specific simulation workflows. Basilisk has been applied to spacecraft dynamics analysis, mission concept development, flight software testing, and autonomy research~\cite{kenneally2020basilisk}. A simplified representation of the Basilisk execution hierarchy and module-message architecture is shown in Fig.~\ref{fig:basilisk_architecture}.

\subsection{Vizard}
\begin{figure}[htbp]
    \centering
    \includegraphics[width=0.8\linewidth]{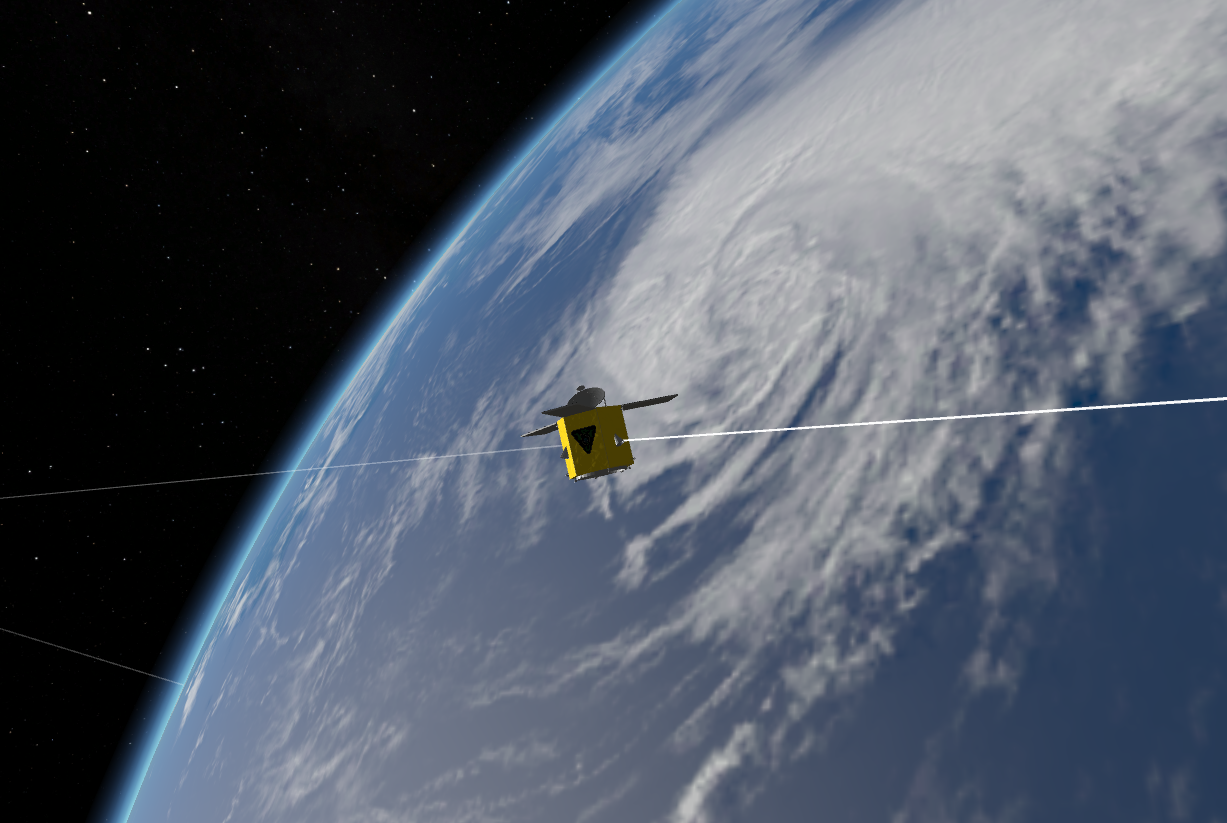}
    \caption{Example spacecraft visualization rendered using 
             Vizard during an Earth-orbit Basilisk simulation.}
    \label{fig:viz-attitude-control}
\end{figure}

Vizard is a Unity-based three-dimensional visualization application developed alongside Basilisk for interactive rendering of simulation environments and spacecraft states~\cite{wood2018flexible}. The visualization framework interfaces directly with Basilisk simulations and supports both live streaming and offline playback workflows. A key characteristic of Vizard is that it displays only elements explicitly modeled in the simulation --- components absent from the simulation model do not appear in the visualization, providing a useful diagnostic aid for identifying missing simulation elements. Vizard may be used to visualize spacecraft trajectories, reference frames, actuator states, celestial bodies, and simulation telemetry within a unified graphical environment. An example visualization is shown in Fig.~\ref{fig:viz-attitude-control}.

\subsection{Docker and Docker Compose}

Docker is an open-source containerization platform that enables applications and their dependencies to be packaged into portable execution environments~\cite{merkel2014docker}. In scientific and engineering workflows, containerization simplifies dependency management, reduces platform-specific configuration issues, and improves portability across heterogeneous development systems~\cite{rad2017introduction, boettiger2015introduction}. Docker Compose extends this capability by providing a declarative configuration mechanism through a \texttt{docker-compose.yml} file, enabling container services, runtime properties, and volume mounts to be defined and launched with a single command~\cite{hasan2021study}.

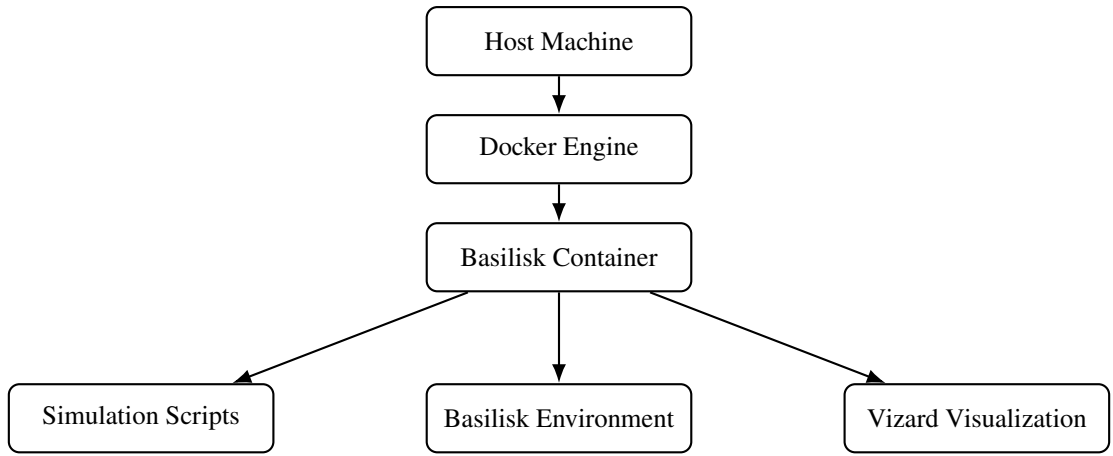
\begin{figure}[htbp]
\centering
\begin{tikzpicture}[
    node distance=0.5cm,
    every node/.style={font=\small},
    block/.style={
        rectangle, rounded corners, draw=black, thick,
        minimum width=3.5cm, minimum height=0.9cm, align=center
    },
    arrow/.style={-{Latex[length=2.5mm]}, thick}
]
\node[block] (host)      {Host Machine};
\node[block, below=of host]      (docker)    {Docker Engine};
\node[block, below=of docker]    (container) {Basilisk Container};
\node[block, below left=1.2cm and 2.0cm of container]  
    (scripts)  {Simulation Scripts};
\node[block, below=1.2cm of container]                 
    (basilisk) {Basilisk Environment};
\node[block, below right=1.2cm and 2.0cm of container] 
    (viz)      {Vizard Visualization};
\draw[arrow] (host)      -- (docker);
\draw[arrow] (docker)    -- (container);
\draw[arrow] (container) -- (scripts);
\draw[arrow] (container) -- (basilisk);
\draw[arrow] (container) -- (viz);
\end{tikzpicture}
\caption{Simplified containerized Basilisk workflow using 
         Docker and Docker Compose.}
\label{fig:docker_workflow}
\end{figure}

Within the workflow presented in this work, Docker is used to encapsulate the Basilisk build environment, supporting libraries, and simulation scripts within a consistent runtime configuration. Environment construction is defined through a \texttt{Dockerfile}, while container execution and runtime settings are managed through Docker Compose. The overall containerized deployment workflow is summarized in Fig.~\ref{fig:docker_workflow}.

\section{Containerized Basilisk Environment}
\label{sec:containerization}

\subsection{Docker-Based Deployment}
The workflow presented in this work is based on Ubuntu~22.04 and Python~3.10, which are compatible with the current Basilisk build requirements~\cite{kenneally2020basilisk}. Environment construction is defined through a \texttt{Dockerfile} containing the complete build configuration, while runtime services and container settings are managed through a \texttt{docker-compose.yml} configuration file. The containerized environment installs Basilisk dependencies, configures the Python environment, and builds the Basilisk framework from source, including scientific Python libraries for simulation analysis and post-processing. The containerized environment builds Basilisk from the repository state corresponding to commit \href{https://github.com/AVSLab/basilisk/commit/87cb4116e09694ddf6587f61a7f4196d2e720c43}{87cb4116e09694ddf6587f61a7f4196d2e720c43}, ensuring reproducibility across future changes to the upstream develop branch.

\subsection{Directory Structure and Bind Mounts}
The containerized workflow utilizes bind mounts to synchronize directories between the host system and the Docker container, enabling simulation scripts and configuration files edited on the host to remain immediately accessible within the containerized environment~\cite{hasan2021study}. The recommended project directory structure is:

\vspace{0.5cm}
\begin{minipage}{0.3\textwidth}
    \scriptsize
    Inside Docker Container\newline
    \dirtree{%
        .1 /app.
        .2 basilisk.
        .3 conanfile.py.
        .3 \textcolor{orange}{dist3}.
        .3 setup.py.
        .3 src.
        .3 supportData.
        .3 :.
        .2 workspace.
    }
\end{minipage}
\vrule
\hfill 
\begin{minipage}{0.3\textwidth}
    \scriptsize
    Inside Docker Container\newline
    \dirtree{%
        .1 /app.
        .2 basilisk.
        .2 \textcolor{red}{workspace}.
        .3 \textcolor{cyan}{docker}.
        .4 \textcolor{cyan}{Dockerfile}.
        .3 \textcolor{cyan}{.env}.
        .3 \textcolor{cyan}{docker-compose.yml}.
        .3 \textcolor{cyan}{build-basilisk.sh}.
        .3 \textcolor{cyan}{:}.
    }
\end{minipage}
\vrule
\hfill 
\begin{minipage}{0.3\textwidth}
    \scriptsize
    Structure on Windows (Host)\newline\newline
    \dirtree{%
        .1 /Local Drive.
        .2 \textcolor{red}{basilisk-docker}.
        .3 \textcolor{cyan}{docker}.
        .4 \textcolor{cyan}{Dockerfile}.
        .3 \textcolor{cyan}{.env}.
        .3 \textcolor{cyan}{docker-compose.yml}.
        .3 \textcolor{cyan}{build-basilisk.sh}.
        .3 \textcolor{cyan}{:}.
    }
\end{minipage}
\vspace{0.5cm}

Inside the container, the working directory is \texttt{/app}, under which Basilisk resides alongside the bind-mounted user directory.

\subsection{Dockerfile}
The \texttt{Dockerfile} defines the complete image build sequence. Basilisk is compatible with Ubuntu~22.04 and Python~3.10. The \texttt{DEBIAN\_FRONTEND=noninteractive} environment variable suppresses interactive prompts during package installation, with \texttt{TZ} set to configure the container timezone. The \texttt{git-lfs} package is required to recover large support files exceeding 100~MB that were stripped during the migration of Basilisk from Bitbucket to GitHub (AVSLab/basilisk discussion~\#60, \url{https://github.com/AVSLab/basilisk/discussions/60}). Basilisk is built using a single invocation of \texttt{conanfile.py} with both \texttt{--buildProject False} and \texttt{--vizInterface True} flags specified together, the latter enabling the Vizard interface. The packages are pinned to a compatible version range consistent with current Basilisk build requirements.

\begin{lstlisting}[language=bash]
FROM ubuntu:22.04

ARG BASILISK_COMMIT=87cb4116e09694ddf6587f61a7f4196d2e720c43
LABEL basilisk_commit=${BASILISK_COMMIT}

ENV DEBIAN_FRONTEND=noninteractive \
    TZ=America/Los_Angeles

RUN apt update && \
    apt-get install -y \
    git git-lfs build-essential \
    python3-setuptools python3-dev \
    python3-tk python3-pip python3-venv \
    swig libgtk2.0-0 libzmq3-dev \
    && rm -rf /var/lib/apt/lists/*

WORKDIR /app

RUN git clone https://github.com/AVSLab/basilisk.git && \
    cd basilisk && \
    git checkout ${BASILISK_COMMIT} && \
    git lfs pull && \
    python3 -m pip install --upgrade pip && \
    python3 -m pip install --no-cache-dir \
        "conan>=2.0.5,<=2.15.1" \
        "cmake>=3.26,<4.0" \
        "setuptools>=70.1.0,<=78.1.0" \
        "wheel>=0.45.1,<=0.46.1" \
        "setuptools-scm>=8.0,<=8.2.1" \
        "numpy>=1.24.4,<2.4.0" \
        "pandas>=2.0.3,<=2.3.3" \
        "matplotlib>=3.7.5,<=3.10.7" \
        "bokeh>=3.4.0,<=3.8.0" \
        "scipy>=1.10.1,<2.0" \
        "pyyaml>=6.0,<7.0" \
        "pillow>=10.4.0,<=12.0.0" \
        "requests>=2.32.3,<=2.32.5" \
        "packaging>=24,<26" \
        "tqdm==4.67.1" \
        "pytest>=8.3.5,<=9.0.1" \
        "pytest-html==4.1.1" \
        "pytest-xdist>=3.6.1,<=3.8.0" \
        "pytest-timeout==2.4.0" \
        "pytest-rerunfailures>=13.0,<=16.1" \
        "pre-commit>=3.5.0,<=4.5.0" \
        "clang-format>=20.1.0,<=21.1.6" \
        "psutil>=7.0.0,<=7.1.3" \
        "swig>=4.4.1,<5" \
        "protobuf>=5.29.4,<=6.33.1" \
        "libclang>=15.0.6.1,<=18.1.1" \
        "colorama" \
        "ipykernel" && \
    python3 conanfile.py \
        --buildProject False \
        --vizInterface True \
        --managePipEnvironment False && \
    cd dist3 && \
    make -j"$(nproc)"

WORKDIR /app/workspace
\end{lstlisting}

\subsection{Docker Compose Configuration}
The \texttt{docker-compose.yml} file specifies container runtime properties including the bind mount, working directory, and environment file reference:

\begin{lstlisting}[language=HTML]
services:
  bsk:
    container_name: basilisk_gnc
    image: bsk_img
    build:
      context: ./docker
    working_dir: /app/workspace
    env_file:
      - .env
    volumes:
      - type: bind
        source: .
        target: /app/workspace
    tty: true
    stdin_open: true
\end{lstlisting}

The \texttt{tty} and \texttt{stdin\_open} flags allocate a pseudo-TTY and keep standard input open, enabling interactive terminal sessions within the running container. The bind mount maps the host project directory to \texttt{/app/workspace} inside the container, allowing bidirectional access to simulation scripts without file copying~\cite{hasan2021study}.

\subsection{Environment File}
The \texttt{.env} file configures the Python path and suppresses bytecode generation:

\begin{lstlisting}[language=bash]
PYTHONPATH=/app/basilisk/dist3
PYTHONDONTWRITEBYTECODE=1
PYTHONUNBUFFERED=1
\end{lstlisting}

The \texttt{PYTHONPATH} variable ensures the Python interpreter locates compiled Basilisk modules without requiring explicit path manipulation in each simulation script.

\subsection{Building and Launching the Environment}
With the configuration files in place, the container is built and started with a single command:

\begin{lstlisting}[language=bash]
docker compose up -d --build
\end{lstlisting}

The \texttt{--build} flag ensures any changes to the \texttt{Dockerfile} are incorporated before starting the container. The \texttt{-d} flag launches services in detached mode, running the container in the background without blocking the terminal session.

\section{Basilisk Architecture}
\label{sec:architecture}

This section describes the primary software architecture concepts underlying Basilisk simulations, including the module-message system, the process-task execution hierarchy, the module lifecycle, and mechanisms for data exchange and recording. These concepts form the foundation for the simulation workflows presented in Sections~\ref{sec:standalone} and~\ref{sec:bsksim}.

\subsection{Modules and Messages}
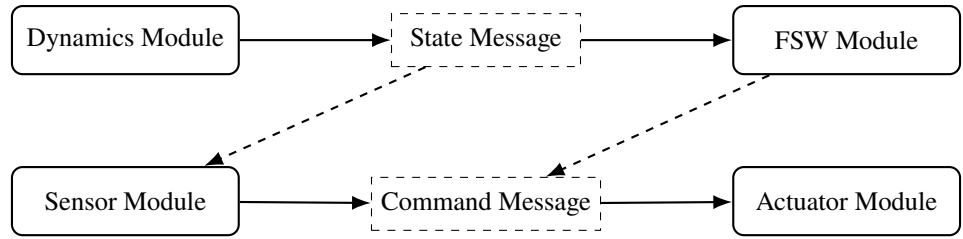
\begin{figure}[htbp]
\centering

\begin{tikzpicture}[
    node distance=2.0cm,
    every node/.style={font=\small},
    module/.style={
        rectangle,
        rounded corners,
        draw=black,
        thick,
        minimum width=3.0cm,
        minimum height=0.9cm,
        align=center
    },
    msg/.style={
        rectangle,
        draw=black,
        dashed,
        minimum width=2.5cm,
        minimum height=0.7cm,
        align=center
    },
    arrow/.style={
        -{Latex[length=2.5mm]},
        thick
    }
]

\node[module] (dyn) {Dynamics Module};
\node[msg, right=of dyn] (state) {State Message};
\node[module, right=of state] (fsw) {FSW Module};
\node[msg, below=1.45cm of state] (cmd) {Command Message};
\node[module, below=1.2cm of dyn] (sensor) {Sensor Module};
\node[module, below=1.2cm of fsw] (actuator) {Actuator Module};

\draw[arrow] (dyn) -- (state);
\draw[arrow] (state) -- (fsw);
\draw[arrow] (sensor) -- (cmd);
\draw[arrow] (cmd) -- (actuator);
\draw[arrow,dashed] (fsw) -- (cmd);
\draw[arrow,dashed] (state) -- (sensor);

\end{tikzpicture}

\caption{Representative Basilisk module-message architecture.}
\label{fig:module_message_architecture}

\end{figure}

The fundamental building blocks of a Basilisk simulation are modules and messages. A module encapsulates a discrete simulation component --- such as a spacecraft dynamics model, sensor, estimation routine, guidance algorithm, or control law --- and exposes typed input and output message interfaces through which simulation states and control data propagate without direct coupling between module implementations~\cite{kenneally2020basilisk}. Basilisk supports C, C++, and Python module implementations, with compiled C and C++ modules preferred for performance-critical components. The message-passing architecture enables modular construction of simulation workflows while maintaining separation between spacecraft dynamics, flight software, sensor models, and visualization utilities. Figure~\ref{fig:module_message_architecture} illustrates a representative module-message interaction within a Basilisk simulation.

\subsection{Process and Task Hierarchy}
Module execution in Basilisk is organized through a two-level hierarchy of processes and tasks \cite{kenneally2020basilisk}. A task is a named execution group with a fixed update rate, specified in nanoseconds, to which one or more modules are assigned. A process is a named collection of related tasks. The simulation container, created from \texttt{SimulationBaseClass.SimBaseClass()}, manages all processes and controls their execution.

The resulting hierarchy --- simulation container $\rightarrow$ process $\rightarrow$ task $\rightarrow$ module --- determines the execution ordering and timing behavior of the simulation. Modules assigned to the same task execute at a common update rate, while tasks and processes may themselves be prioritized relative to one another. Optional integer priorities may additionally be assigned to modules within a task to control execution ordering.

The following example illustrates the creation of dynamics and flight software processes, each containing tasks at specified update rates:

\begin{lstlisting}[language=Python]
sim_obj = SimulationBaseClass.SimBaseClass()
dyn_process = sim_obj.CreateNewProcess("dynamicsProcess")
dyn_task = sim_obj.CreateNewTask("dynamicsTask", macros.sec2nano(5.))
dyn_process.addTask(dyn_task)
module_obj = cModuleTemplate.cModuleTemplate()
sim_obj.AddModelToTask("dynamicsTask", module_obj, 10)
\end{lstlisting}

\begin{figure}[htbp]
\centering

\begin{tikzpicture}[
    every node/.style={font=\small},
    block/.style={
        rectangle,
        rounded corners,
        draw=black,
        thick,
        minimum width=3.3cm,
        minimum height=0.9cm,
        align=center
    },
    arrow/.style={
        thick,
        -{Latex[length=2.5mm]}
    }
]

\node[block] (sim) at (0,0)
{Simulation Container};

\node[block] (dynproc) at (-4,-2)
{Dynamics Process};

\node[block] (fswproc) at (4,-2)
{FSW Process};

\node[block] (dyntask) at (-4,-4)
{Dynamics Task};

\node[block] (fswtask) at (4,-4)
{FSW Task};

\node[block] (sc) at (-6,-6)
{Spacecraft Module};

\node[block] (sensor) at (-2,-6)
{Sensor Module};

\node[block] (guidance) at (2,-6)
{Guidance Module};

\node[block] (control) at (6,-6)
{Control Module};

\draw[arrow] (sim) -- (dynproc);
\draw[arrow] (sim) -- (fswproc);

\draw[arrow] (dynproc) -- (dyntask);
\draw[arrow] (fswproc) -- (fswtask);

\draw[arrow] (dyntask) -- (sc);
\draw[arrow] (dyntask) -- (sensor);

\draw[arrow] (fswtask) -- (guidance);
\draw[arrow] (fswtask) -- (control);

\end{tikzpicture}

\caption{Representative Basilisk execution hierarchy consisting of processes, tasks, and simulation modules.}
\label{fig:process_task_hierarchy}

\end{figure}
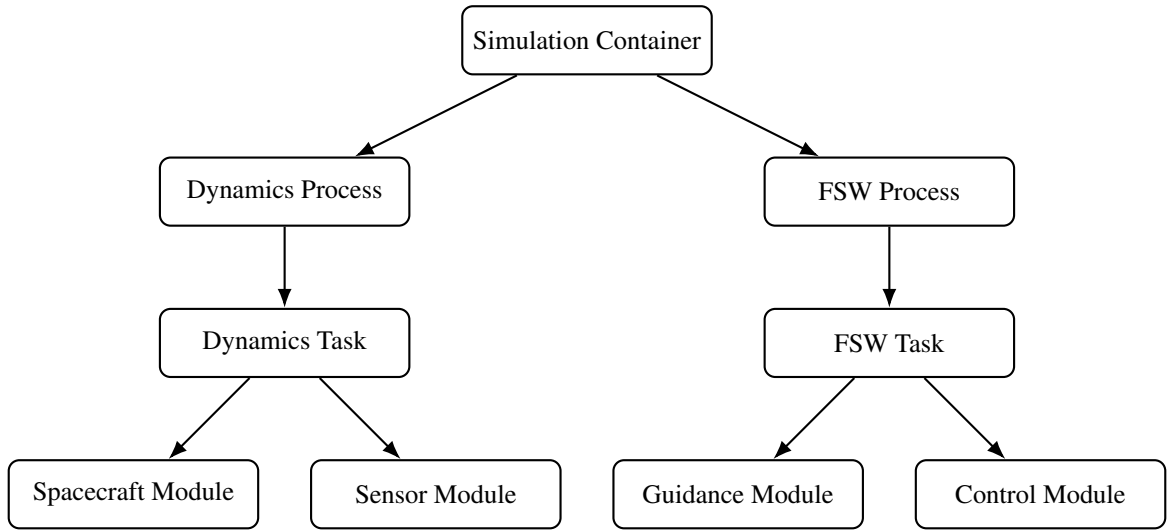

The execution order of modules within a task can be inspected at runtime using \texttt{ShowExecutionOrder()}, or visualized as a figure using \texttt{ShowExecutionFigure(True)} of the \texttt{sim\_obj}. A representative Basilisk execution hierarchy is shown in Fig.~\ref{fig:process_task_hierarchy}.

\subsection{Module Execution Lifecycle}
Simulation modules in Basilisk follow a staged execution lifecycle consisting of three phases~\cite{kenneally2020basilisk}. \texttt{SelfInit()} is called during initialization for C modules, configuring output message connections. \texttt{Reset()} is called by \texttt{InitializeSimulation()} and synchronizes module state to desired defaults, making it the appropriate location for repeatable initialization logic. \texttt{Update()} is called at every task time step and performs the module's primary computation. C++ modules override \texttt{Reset()} and \texttt{UpdateState()} inherited from the \texttt{SysModel} base class.

The following example demonstrates lifecycle behavior, showing how a module variable evolves across initialization and single-step execution:

\begin{lstlisting}[language=Python]
module_obj.dummy = -10
print(f'Before initialization: {module_obj.dummy}')  # -10.0

sim_obj.InitializeSimulation()
print(f'After initialization: {module_obj.dummy}')   # 0.0

sim_obj.TotalSim.SingleStepProcesses()
print(f'After execution: {module_obj.dummy}')        # 1.0
\end{lstlisting}

The \texttt{SingleStepProcesses()} method advances the simulation by a single time step and is particularly useful for testing module input-output behavior in isolation.

\subsection{Message Connection and Recording}
Modules are connected by subscribing an input message to an output message of another module:

\begin{lstlisting}[language=Python]
module_b.dataInMsg.subscribeTo(module_a.dataOutMsg)
\end{lstlisting}

Simulation data is recorded by attaching a recorder object to a task prior to calling the function \texttt{InitializeSimulation()}. Recorders may capture all updates or sample at a specified period:

\begin{lstlisting}[language=Python]
# Record every update
msg_log = module_obj.dataOutMsg.recorder()
sim_obj.AddModelToTask("dynamicsTask", msg_log)

# Record at specified interval
msg_log_sparse = module_obj.dataOutMsg.recorder( macros.sec2nano(20.) )
sim_obj.AddModelToTask("dynamicsTask", msg_log_sparse)
\end{lstlisting}

Logged data is accessible as NumPy arrays after simulation execution, with simulation time retrievable via the \texttt{times()} method on the recorder object.

\section{Standalone-Script Simulation Workflows}
\label{sec:standalone}

Standalone Basilisk simulations are implemented through Python scenario scripts that configure the simulation container, define execution processes and tasks, instantiate spacecraft and environment modules, and execute the simulation timeline. This section presents three representative standalone scenarios of increasing complexity, followed by data logging and visualization workflows. The code snippets show selected lines; full scripts available at \cite{gupta2026basiliskdockerrepo}.

\subsection{Scenario Configuration}
As a recommended coding practice, mission-specific parameters should be externalized from simulation scripts into configuration files. A YAML-based configuration approach separates simulation and spacecraft parameters from implementation logic, improving reusability, and reducing hard-coded values:

\begin{lstlisting}[language=HTML]
simulation:
  - simulation_process_name: simulation_process
  - simulation_task_name:    simulation_task
  - simulation_time:         1000.0
  - simulation_time_unit:    sec
  - time_step:               1.0
spacecraft:
  - mass:    750.0
  - inertia: [900.0, 0.0, 0.0,
               0.0, 800.0, 0.0,
               0.0,   0.0, 700.0]
  - name:    bsk_sat
\end{lstlisting}

A physical consistency check on the inertia matrix is recommended before simulation: the sum of any two diagonal elements must exceed the third, i.e., $I_i + I_j \geq I_k$ for $i \neq j \neq k$. A spacecraft defined by a physically inconsistent inertia matrix will not achieve stable attitude control.

\subsection{Spacecraft in Simulation}
The simplest standalone scenario creates a spacecraft object within the Basilisk simulation container without specifying orbital initial conditions. This provides a useful baseline for verifying the simulation infrastructure and Vizard connectivity before introducing gravitational bodies or control algorithms. The general simulation script structure follows the pattern shown in Fig.~\ref{fig:standalone_flowchart}:

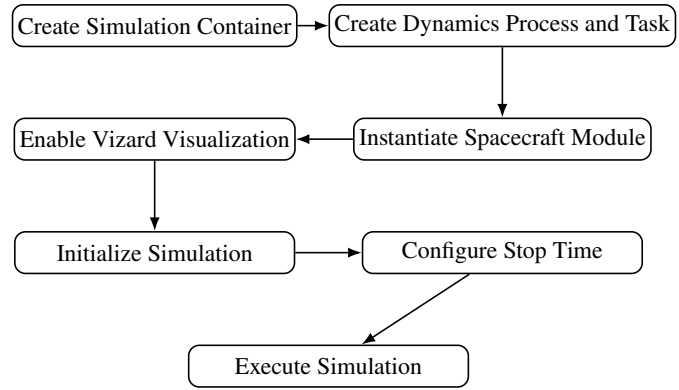
\begin{figure}[htbp]
\centering

\begin{tikzpicture}[
    every node/.style={font=\footnotesize},
    block/.style={
        rectangle,
        rounded corners,
        draw=black,
        line width=0.6pt,
        minimum width=3.7cm,
        minimum height=0.55cm,
        align=center,
        inner sep=2pt
    },
    arrow/.style={
        -{Latex[length=1.8mm]},
        line width=0.6pt
    }
]

\node[block] (a) at (0,0)
{Create Simulation Container};

\node[block] (b) at (4.6,0)
{Create Dynamics Process and Task};

\node[block] (c) at (4.6,-1.5)
{Instantiate Spacecraft Module};

\node[block] (d) at (0,-1.5)
{Enable Vizard Visualization};

\node[block] (e) at (0,-3.0)
{Initialize Simulation};

\node[block] (f) at (4.6,-3.0)
{Configure Stop Time};

\node[block] (g) at (2.3,-4.5)
{Execute Simulation};

\draw[arrow] (a) -- (b);
\draw[arrow] (b) -- (c);
\draw[arrow] (c) -- (d);
\draw[arrow] (d) -- (e);
\draw[arrow] (e) -- (f);
\draw[arrow] (f) -- (g);

\end{tikzpicture}

\caption{Standalone Basilisk scenario execution workflow.}
\label{fig:standalone_flowchart}

\end{figure}

\begin{lstlisting}[language=Python]
from Basilisk.simulation import spacecraft
from Basilisk.utilities import SimulationBaseClass, macros, vizSupport

def run(time_step, simulation_time):
    simulation_obj = SimulationBaseClass.SimBaseClass()
    dynamics_process = simulation_obj.CreateNewProcess("simulation_process")
    simulation_time_step = macros.sec2nano(time_step)
    dynamics_process.addTask(
        simulation_obj.CreateNewTask(
            "simulation_task", simulation_time_step))

    spacecraft_obj = spacecraft.Spacecraft()
    spacecraft_obj.ModelTag = "bsk_sat"
    simulation_obj.AddModelToTask("simulation_task", spacecraft_obj)

    vizSupport.enableUnityVisualization(
        simulation_obj, "simulation_task",
        spacecraft_obj, liveStream=False,
        saveFile=__file__)

    simulation_obj.InitializeSimulation()
    simulation_obj.ConfigureStopTime( macros.sec2nano(simulation_time) )
    simulation_obj.ExecuteSimulation()

if __name__ == "__main__":
    run(1.0, 1000.0)
\end{lstlisting}

A critical step that is frequently overlooked is the call to \texttt{AddModelToTask()}. Omitting this call does not raise an error but results in a simulation that executes without propagating any dynamics. The \texttt{ModelTag} field uniquely identifies the module and is particularly useful when simulating multiple spacecraft.

\begin{figure}[htbp]
    \centering
    \includegraphics[width=0.85\linewidth]{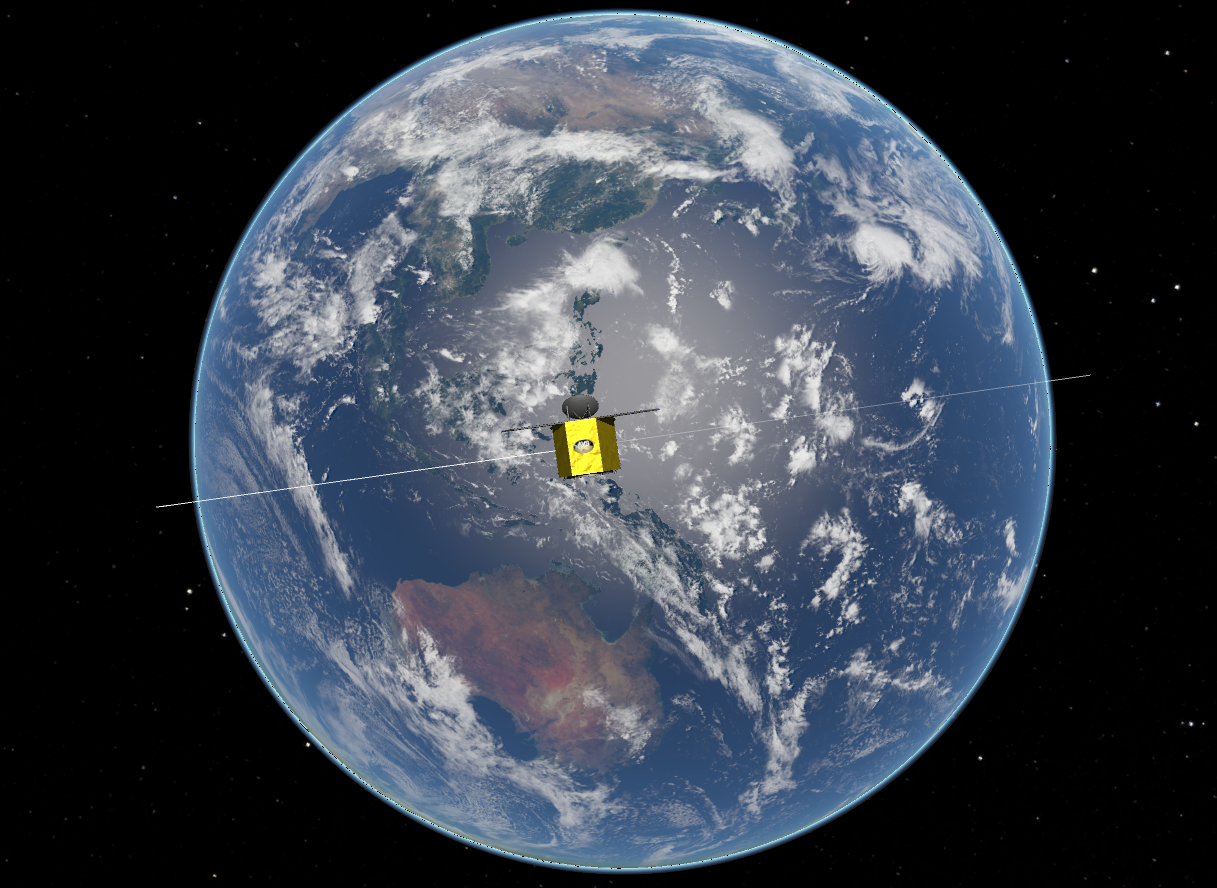}
    \caption{Spacecraft in low-Earth orbit rendered 
             in Vizard from the Earth-orbit scenario.}
    \label{fig:viz-earth-orbit}
\end{figure}

\subsection{Spacecraft in Earth Orbit}
The Earth-orbit scenario extends the base script by introducing a gravitational body and Keplerian initial conditions. Earth is created as the central body using \texttt{gravBodyFactory}, with optional $J_2$ spherical harmonic perturbations. Orbital initial conditions are specified as classical Keplerian elements and converted to inertial position and velocity vectors using \texttt{elem2rv()}:

\begin{lstlisting}[language=Python]
from Basilisk.utilities import simIncludeGravBody, orbitalMotion
import numpy as np

grav_factory = simIncludeGravBody.gravBodyFactory()
planet = grav_factory.createEarth()
planet.isCentralBody = True

if use_spherical_harmonics:
    planet.useSphericalHarmonicsGravityModel(\
        bskPath + '/supportData/LocalGravData/GGM03S-J2-only.txt', 2)

mu = planet.mu
spacecraft_obj.gravField.gravBodies = \
    spacecraft.GravBodyVector(
        list(grav_factory.gravBodies.values()))

oe = orbitalMotion.ClassicElements()
oe.a     = 7000. * 1000    # m
oe.e     = 0.0001
oe.i     = 33.3  * macros.D2R
oe.Omega = 48.2  * macros.D2R
oe.omega = 347.8 * macros.D2R
oe.f     = 85.3  * macros.D2R

r_N, v_N = orbitalMotion.elem2rv(mu, oe)
spacecraft_obj.hub.r_CN_NInit = r_N
spacecraft_obj.hub.v_CN_NInit = v_N

n = np.sqrt(mu / oe.a**3)
T = 2. * np.pi / n
\end{lstlisting}

The orbital period $T$ is computed from the mean motion $n$ as:
\begin{equation}
n = \sqrt{\frac{\mu}{a^3}}, \qquad
T = \frac{2\pi}{n}
\label{eq:orbital_period}
\end{equation}
The simulation duration is a configurable parameter, set to 1000 s in the default invocation shown above. The orbital period $T$ is computed from the mean motion for reference. When spherical harmonics are enabled, a longer duration -- on the order of $3 T \approx 18000$ s for this orbit -- is recommended to observe $J_2$-induced nodal precession. The GGM03S gravity model, derived from four years of GRACE observations between 2003 and 2006, provides the $J_2$ perturbation data~\cite{tapley2007ggm03}. The non-Keplerian equations of motion including the $J_2$ perturbation acceleration $\mathbf{a}_{J_2}$ are given by~\cite{battin1999introduction, bate2020fundamentals}:
\begin{equation}
\ddot{\mathbf{r}} = -\frac{\mu}{r^2}\hat{\mathbf{r}} 
                    + \mathbf{a}_{J_2}
\label{eq:eom_j2}
\end{equation}

The resulting simulation renders the spacecraft in a low-Earth orbit configuration within Vizard, as shown in Fig.~\ref{fig:viz-earth-orbit}.
 
\subsection{Sun-Earth System with SPICE Ephemerides}
The Sun-Earth scenario introduces a second gravitational body and incorporates planetary ephemeris data through the SPICE toolkit~\cite{acton1998overview}. SPICE (Spacecraft, Planet, Instrument, Camera-matrix, Events) is an information system developed by NASA's Navigation and Ancillary Information Facility (NAIF) that provides observation geometry --- positions, velocities, and orientations of spacecraft and celestial bodies --- through data files called kernels~\cite{acton2011spice, costa2018spice}. The required kernels for this simulation are the solar system ephemeris (\texttt{de430.bsp}), leap second file (\texttt{naif0012.tls}), solar system masses (\texttt{de-403-masses.tpc}), and planetary constants (\texttt{pck00010.tpc}).

\begin{lstlisting}[language=Python]
from Basilisk.simulation import ephemerisConverter
from Basilisk.topLevelModules import pyswice

grav_bodies = grav_factory.createBodies(['sun', 'earth'])
grav_bodies['earth'].isCentralBody = True

sun = 0     # Assign identity to both celestial bodies; sun is unused in this code
earth = 1
    
time_init = "2000 Jan 1 11:59:28.000 (UTC)"
spacecraft_obj.gravField.gravBodies = \
    spacecraft.GravBodyVector(
        list(grav_factory.gravBodies.values()))
grav_factory.createSpiceInterface(
    bskPath + '/supportData/EphemerisData/',
    time_init, epochInMsg=True)
epoch_msg = grav_factory.epochMsg
simulation_obj.AddModelToTask("simulation_task", grav_factory.spiceObject)

earth_ephem = ephemerisConverter.EphemerisConverter()
simulation_obj.AddModelToTask("simulation_task", earth_ephem)

grav_factory.spiceObject.zeroBase = 'Earth'
earth_ephem.addSpiceInputMsg(grav_factory.spiceObject.planetStateOutMsgs[earth])

pyswice.furnsh_c(grav_factory.spiceObject.SPICEDataPath + 'de430.bsp')
pyswice.furnsh_c(grav_factory.spiceObject.SPICEDataPath + 'naif0012.tls')
pyswice.furnsh_c(grav_factory.spiceObject.SPICEDataPath + 'de-403-masses.tpc')
pyswice.furnsh_c(grav_factory.spiceObject.SPICEDataPath + 'pck00010.tpc')

viz.epochInMsg.subscribeTo(epoch_msg)
\end{lstlisting}

\begin{figure}[htbp]
    \centering
    \includegraphics[width=0.8\linewidth]{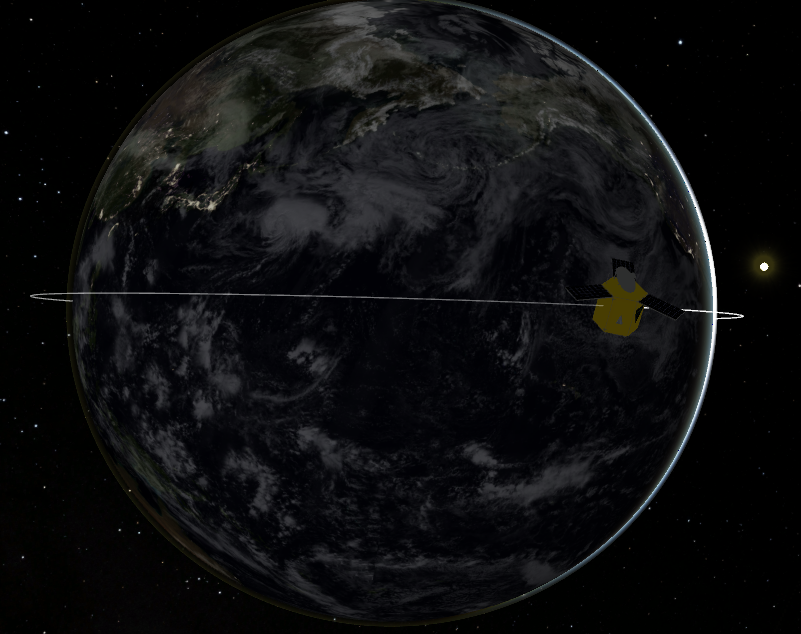}
    \caption{Sun-Earth-spacecraft simulation rendered 
             in Vizard using SPICE ephemerides.}
    \label{fig:sun-earth-vizard}
\end{figure}

The Ephemeris DE430 models the damping term between the Moon's liquid core and solid mantle, making it suitable for epochs between 1550 and 2650~\cite{folkner2014planetary}. The \texttt{zeroBase} field is set to \texttt{Earth} to express all state vectors relative to Earth's center, consistent with the Earth-centered orbit configuration. An example Sun-Earth-spacecraft visualization rendered through Vizard using SPICE ephemerides is shown in Fig.~\ref{fig:sun-earth-vizard}.

\subsection{Data Logging and Visualization}
Spacecraft state data is recorded using the message recorder interface described in Section~\ref{sec:architecture}. The sampling period is computed from the total simulation time, time step, and desired number of data points:
\begin{equation}
\Delta t_\text{samp} = 
    \left\lfloor \frac{T_{\text{final}}}{\Delta t_\text{sim} \cdot (N-1)} 
    \right\rfloor \Delta t_\text{sim}
\label{eq:sampling}
\end{equation}
where $T_{\text{final}}$ is the total simulation time and $N$ is the number of data points. Basilisk enforces a minimum sampling period of 1~nanosecond.

\begin{lstlisting}[language=Python]
sampling_time = unitTestSupport.samplingTime(
    simulation_time, simulation_time_step, 
    num_data_points)

sc_log = spacecraft_obj.scStateOutMsg.recorder(sampling_time)
simulation_obj.AddModelToTask("simulation_task", sc_log)

simulation_obj.InitializeSimulation()
simulation_obj.ConfigureStopTime(simulation_time)
simulation_obj.ExecuteSimulation()

r_BN_N = sc_log.r_BN_N
v_BN_N = sc_log.v_BN_N
t_s    = sc_log.times() * macros.NANO2SEC
\end{lstlisting}

Position, velocity, and orbital element histories are exportable to CSV for post-processing. The \texttt{rv2elem()} function converts logged inertial state vectors back to classical orbital elements at each time step. Visualization through Vizard provides interactive three-dimensional rendering of spacecraft trajectories, orbital geometry, and celestial body configurations, as shown in Figs.~\ref{fig:viz-earth-orbit} and~\ref{fig:sun-earth-vizard}.

\section{BSKSim-Based Simulation Workflows}
\label{sec:bsksim}

The standalone scripting approach presented in Section~\ref{sec:standalone} is well-suited for simple scenarios but becomes difficult to maintain as simulation complexity grows. The BSKSim framework addresses this by providing an object-oriented architecture that separates dynamics models, flight software models, and scenario configuration into distinct, reusable components. This section describes the BSKSim directory structure, class hierarchy, dynamics and flight software implementations, and Monte Carlo simulation capability.

\subsection{Directory Structure}
The BSKSim workflow organizes simulation components into a structured directory hierarchy:

\vspace{0.5cm}
\begin{minipage}{\textwidth}
    \scriptsize
    \dirtree{%
        .1 /workspace.
        .2 models/.
        .3 BSK\_Dynamics.py.
        .3 BSK\_Fsw.py.
        .2 scenarios/.
        .3 scenario\_basic\_orbit.py.
        .3 scenario\_attitude\_control.py.
        .3 scenario\_monte\_carlo.py.
        .2 utilities/.
        .3 BSK\_masters.py.
        .3 BSK\_Plotting.py.
    }
\end{minipage}
\vspace{0.5cm}

The \texttt{models} directory contains dynamics and flight software class definitions. The \texttt{scenarios} directory contains scenario-specific scripts that inherit from the master classes. The \texttt{utilities} directory contains the master class definitions and shared plotting utilities.

\subsection{Class Hierarchy}
The BSKSim architecture is built around two base classes defined in \texttt{BSK\_masters.py}: \texttt{BSKSim}, which inherits from \texttt{SimulationBaseClass.SimBaseClass()} and manages process and model registration, and \texttt{BSKScenario}, which defines the interface for scenario-specific initialization, logging, and output extraction. A scenario class inherits from both:

\begin{lstlisting}[language=Python]
class AttitudeControl(BSKSim, BSKScenario):
    def __init__(self):
        super(AttitudeControl, self).__init__()
        self.set_dynamics_model(BSK_Dynamics)
        self.set_fsw_model(BSK_Fsw)
        self.configure_initial_conditions()
        self.log_outputs()

    def configure_initial_conditions(self): ...
    def log_outputs(self):                  ...
    def pull_outputs(self, show_plots):     ...
\end{lstlisting}

The \texttt{BSKSim} base class exposes \texttt{set\_dynamics\_model()} and \texttt{set\_fsw\_model()} methods that instantiate the corresponding model classes and register their processes. The \texttt{BSKScenario} base class defines \texttt{configure\_initial\_conditions()}, \texttt{log\_outputs()}, and \texttt{pull\_outputs()} as abstract interfaces that each scenario overrides. This separation ensures that dynamics and flight software implementations remain reusable across multiple scenarios without modification.

\subsection{Dynamics Model}
The dynamics model class \texttt{BSKDynamicModels} instantiates and configures all dynamics-related modules, including the spacecraft hub, gravitational bodies, navigation sensors, and external force/torque effectors:

\begin{lstlisting}[language=Python]
class BSKDynamicModels():
    def __init__(self, sim_obj, dynamics_rate):
        self.spacecraft_obj = spacecraft.Spacecraft()
        self.grav_factory = simIncludeGravBody.gravBodyFactory()
        self.simple_nav_obj = simpleNav.SimpleNav()
        self.ext_force_torque = extForceTorque.ExtForceTorque()
        self.earth_ephem = ephemerisConverter.EphemerisConverter()

        self.initialize_all_dynamic_objects()

        sim_obj.AddModelToTask(
            self.task_name, 
            self.spacecraft_obj, None, 201)
        sim_obj.AddModelToTask(
            self.task_name, 
            self.simple_nav_obj, None, 109)
        sim_obj.AddModelToTask(
            self.task_name,
            self.grav_factory.spiceObject, 200)
        sim_obj.AddModelToTask(
            self.task_name, 
            self.ext_force_torque, None, 300)
        sim_obj.AddModelToTask(
            self.task_name, 
            self.earth_ephem, 199)
\end{lstlisting}

Module priority values control execution order within the dynamics task. The spacecraft hub (priority 201) executes before the SPICE object (200), which executes before the Earth ephemeris converter (199), ensuring that gravitational state information is available when the spacecraft dynamics are propagated. The external torque effector is assigned the highest priority (300) within the dynamics task, ensuring that any commanded torques are staged before the spacecraft hub integrates the equations of motion at each time step.

The spacecraft hub is configured with mass and inertia properties, and the \texttt{SimpleNav} module subscribes to the spacecraft state output message to provide navigation estimates to the flight software:

\begin{lstlisting}[language=Python]
def set_spacecraft_hub(self):
    self.spacecraft_obj.ModelTag   = "bskSat"
    self.I_sc = [900., 0., 0.,
                   0., 800., 0.,
                   0., 0., 600.]
    self.spacecraft_obj.hub.mHub   = 750.0
    self.spacecraft_obj.hub.IHubPntBc_B = sp.np2EigenMatrix3d(self.I_sc)

def set_simple_nav_obj(self):
    self.simple_nav_obj.scStateInMsg.subscribeTo(
            self.spacecraft_obj.scStateOutMsg)
\end{lstlisting}

\subsection{Flight Software Model}
The flight software model class \texttt{BSKFswModels} instantiates guidance, navigation, and control algorithm modules and connects them through gateway messages. Gateway messages are C-wrapped message objects that provide a common input point for downstream modules that may be written by multiple upstream sources:

\begin{lstlisting}[language=Python]
def setup_gateway_msgs(self, sim_obj):
    self.cmd_torque_msg = messaging.CmdTorqueBodyMsg_C()
    self.attitude_ref_msg = messaging.AttRefMsg_C()
    self.attitude_guid_msg = messaging.AttGuidMsg_C()
    self.zero_gate_way_msgs()
    sim_obj.dynamics_model\
        .ext_force_torque\
        .cmdTorqueInMsg\
        .subscribeTo(self.cmd_torque_msg)
\end{lstlisting}

The flight software model implements hill-point guidance, inertial pointing, attitude tracking error computation, and MRP feedback control~\cite{schaub2003analytical}. Tasks and modules are organized by function:

\begin{lstlisting}[language=Python]
# Hill-point guidance
self.hill_point_data.transNavInMsg.subscribeTo(
        sim_obj.dynamics_model.simple_nav_obj.transOutMsg)
self.hill_point_data.celBodyInMsg.subscribeTo(
        sim_obj.dynamics_model.earth_ephem.ephemOutMsgs[0])

# MRP feedback control gains
self.mrp_fbcontrol_data.K               = 3.5
self.mrp_fbcontrol_data.Ki              = -1.0  # negative Ki per BSK MRP controller sign convention
self.mrp_fbcontrol_data.P               = 30.0
self.mrp_fbcontrol_data.integralLimit   = 2. / self.mrp_fbcontrol_data.Ki * 0.1  # = -0.2
\end{lstlisting}

The negative \texttt{Ki} value follows Basilisk's MRP feedback controller sign convention; the resulting negative \texttt{integralLimit} is handled correctly by the module's internal integrator clamp. Flight software tasks are disabled by default and activated through simulation events triggered by a \texttt{mode\_request} variable, enabling clean transitions between standby, inertial pointing, and hill-point modes without restarting the simulation.

\subsection{Scenario Execution}
A scenario is instantiated, executed, and post-processed through a standardized four-step pattern:

\begin{lstlisting}[language=Python]
def run(show_plots):
    # 1. Instantiate
    scenario = AttitudeControl()
    # 2. Set mode and execute
    scenario.mode_request = 'hillPoint'
    scenario.InitializeSimulation()
    scenario.ConfigureStopTime( macros.min2nano(10.) )
    scenario.ExecuteSimulation()
    # 3. Extract outputs
    figure_list = scenario.pull_outputs(show_plots)
    return figure_list
\end{lstlisting}

\subsection{Monte Carlo Simulation}
Basilisk provides native Monte Carlo support through the \texttt{MonteCarloController} class, enabling automated execution of simulation ensembles with randomized parameter variations. A Monte Carlo run is configured by specifying a scenario function, the number of runs, and a set of variable dispersions:

\begin{lstlisting}[language=Python]
from Basilisk.utilities.MonteCarlo.Controller import Controller
from Basilisk.utilities.MonteCarlo.Dispersions import UniformDispersion, NormalVectorCartDispersion

mc_controller = Controller()
mc_controller.setSimulationFunction(run_scenario)
mc_controller.setExecutionCount(100)
mc_controller.setArchiveDir("monte_carlo_results")

mc_controller.addDispersion(
    UniformDispersion(
        "spacecraft_obj.hub.mHub", distributionData=[700., 800.]))

mc_controller.addDispersion(
    NormalVectorCartDispersion(
        "spacecraft_obj.hub.r_CN_NInit", mean=r_N, stdDeviation=1000.))

mc_controller.executeSimulations()
\end{lstlisting}

Each Monte Carlo run executes the scenario function with independently sampled parameter values drawn from the specified distributions. Results are archived to the specified directory and accessible for post-processing using standard Python analysis workflows. The classes \texttt{UniformDispersion} and \texttt{NormalVectorCartDispersion} provide uniform and Gaussian parameter sampling respectively; additional distribution types are available in the Basilisk Monte Carlo utilities library.

\section{Discussion}
\label{sec:discussion}

\subsection{Practical Reproducibility Benefits}
The containerized workflow presented in this work addresses several concrete reproducibility challenges encountered in distributed GN\&C simulation development. By encoding the complete Basilisk build configuration as a \texttt{Dockerfile}, the simulation environment becomes version-controllable, auditable, and reconstructable from a single command. Dependency versions, compiler toolchains, and Python library configurations are fixed within the container image, eliminating the environment-specific variability that frequently causes simulation behavior to diverge across development machines~\cite{boettiger2015introduction}.

The bind-mount architecture preserves the separation between the containerized simulation infrastructure and user-developed scenario scripts. Engineers may develop, modify, and version-control simulation scripts on the host machine using familiar tools while executing them within the consistent containerized environment. This separation also simplifies onboarding: new team members achieve a fully configured Basilisk environment with a single \texttt{docker compose up} command rather than following multi-step manual installation procedures that are frequently platform-dependent.

\subsection{Educational and Research Utility}
The presented workflow is well suited for instructional and rapid-prototyping applications. In educational settings, containerized environments reduce software installation complexity and improve consistency across student development systems, allowing instructors to focus on simulation concepts rather than environment configuration. The progression from standalone scripts to BSKSim-based scenarios presented in Sections~\ref{sec:standalone} and~\ref{sec:bsksim} provides a structured introduction to Basilisk's capabilities suitable for graduate-level GN\&C courses and workshop formats.

Within research contexts, the BSKSim class hierarchy supports rapid development of mission-specific simulation scenarios by enabling dynamics and flight software components to be reused across scenarios without modification. The Monte Carlo capability described in Section~\ref{sec:bsksim} further enables systematic uncertainty quantification and sensitivity analysis within the same containerized environment.

\subsection{Limitations and Known Issues}
Several practical limitations are relevant when deploying the presented workflow. Basilisk undergoes active development, and compatibility between framework releases, Python versions, and Conan package management configurations may require periodic \texttt{Dockerfile} updates as the framework evolves. The \texttt{conan} version pin used in the present configuration reflects a known compatibility requirement that may change in future Basilisk releases.

Simulation workflows utilizing SPICE ephemerides depend on the availability and correct placement of kernel files. The \texttt{de430.bsp} ephemeris file was stripped during the migration of Basilisk from Bitbucket to GitHub due to file size constraints and is recovered through \texttt{git lfs pull} during the container build process as shown in Section~\ref{sec:containerization}. Missing or incorrectly configured kernel files result in simulation initialization failures that may not produce informative error messages.

Graphical visualization through Vizard depends on host-system graphics driver compatibility when using live streaming mode. Offline playback from saved binary files is recommended for containerized workflows where graphics hardware passthrough is unavailable or inconvenient.

Docker images incorporating the full Basilisk build introduce substantial storage overhead due to the compilation of C++ source code and associated dependencies. Build times on the order of 15--20 minutes are typical for first-time image construction, though Docker layer caching significantly reduces rebuild times when only simulation scripts or Python dependencies are modified.

\subsection{Relationship to Companion Works}
This paper accompanies and expands upon the workshop presentation~\cite{gupta2024basilisk}. The containerized deployment approach described here is also relevant to broader reproducibility questions in GN\&C simulation workflows, including cross-platform validation of simulation outputs, which are addressed separately in ongoing work. The scenario configuration and scripting concepts introduced through the BSKSim examples motivate a YAML-driven simulation orchestration framework currently under development as a separate open-source tool, aimed at enabling declarative specification of Basilisk simulation scenarios without direct scripting.

\section{Conclusion}
\label{sec:conclusion}

This paper presented a containerized deployment and simulation workflow for the Basilisk astrodynamics framework using Docker. The workflow encapsulates the complete Basilisk build environment within a portable Docker container, eliminating environment-specific dependency conflicts and enabling consistent simulation execution across heterogeneous development systems. Representative simulation scenarios were presented at increasing levels of complexity, from standalone orbital dynamics scripts to BSKSim-based attitude dynamics and control simulations with Monte Carlo analysis capability.

The Dockerfile, Docker Compose configuration, and example simulation scripts presented in this work provide a self-contained implementation reference for GN\&C engineers and researchers seeking to deploy Basilisk in portable, reproducible simulation environments. The material expands upon the workshop presentation, providing additional implementation detail on the BSKSim class hierarchy, flight software architecture, attitude control implementation, and Monte Carlo simulation workflows not covered in the original presentation~\cite{gupta2024basilisk}.

The complete implementation, including all configuration files and example simulation scripts, is available at \url{https://github.com/theinfinitelabs/basilisk-docker}, see reference~\cite{gupta2026basiliskdockerrepo} for the archived version on Zenodo.

\section*{Acknowledgments}
The author thanks Dr.~Hanspeter Schaub and the AVS Laboratory at the University of Colorado Boulder for developing Basilisk and for assistance in learning the framework. The author also thanks Abhinav Gupta for coding practice guidance and Ishaan Patel for review and feedback on the original workshop presentation. The author acknowledges use of an AI writing assistant for editorial support during manuscript preparation.

\printbibliography

\end{document}